\documentclass[
 reprint,
 amsmath,amssymb,
 aps,
 prl,
]{revtex4-2}

\pdfoutput=1
\usepackage{graphicx}
\usepackage{float}
\usepackage{dcolumn}
\usepackage{bm}
\usepackage{bbm}
\usepackage{braket}
\usepackage{xcolor}
\usepackage[T1]{fontenc}

\makeatletter
\DeclareFontFamily{OMX}{MnSymbolE}{}
\DeclareSymbolFont{MnLargeSymbols}{OMX}{MnSymbolE}{m}{n}
\SetSymbolFont{MnLargeSymbols}{bold}{OMX}{MnSymbolE}{b}{n}
\DeclareFontShape{OMX}{MnSymbolE}{m}{n}{
    <-6>  MnSymbolE5
   <6-7>  MnSymbolE6
   <7-8>  MnSymbolE7
   <8-9>  MnSymbolE8
   <9-10> MnSymbolE9
  <10-12> MnSymbolE10
  <12->   MnSymbolE12
}{}
\DeclareFontShape{OMX}{MnSymbolE}{b}{n}{
    <-6>  MnSymbolE-Bold5
   <6-7>  MnSymbolE-Bold6
   <7-8>  MnSymbolE-Bold7
   <8-9>  MnSymbolE-Bold8
   <9-10> MnSymbolE-Bold9
  <10-12> MnSymbolE-Bold10
  <12->   MnSymbolE-Bold12
}{}

\let\llangle\@undefined
\let\rrangle\@undefined
\DeclareMathDelimiter{\llangle}{\mathopen}%
                     {MnLargeSymbols}{'164}{MnLargeSymbols}{'164}
\DeclareMathDelimiter{\rrangle}{\mathclose}%
                     {MnLargeSymbols}{'171}{MnLargeSymbols}{'171}
\makeatother

\newcommand\identity{1\kern-0.25em\text{l}}

\usepackage{accents}
\newcommand{\dbtilde}[1]{\accentset{\approx}{#1}}

\begin{document}

\title{A photonic which-path entangler  based on  longitudinal cavity-qubit coupling}

\author{Z.~M.~McIntyre }
\email{zoe.mcintyre@mail.mcgill.ca}
\author{W.~A.~Coish}%
 \email{william.coish@mcgill.ca}
\affiliation{%
 Department of Physics, McGill University, 3600 rue University, Montreal, QC, H3A 2T8, Canada
}%

\date{\today}

\begin{abstract}
We show that a modulated longitudinal cavity-qubit coupling can be used to control the path taken by a multiphoton coherent-state wavepacket conditioned on the state of a qubit, resulting in a qubit--which-path (QWP) entangled state. QWP states can generate long-range multipartite entanglement using strategies for interfacing discrete- and continuous-variable degrees-of-freedom. Using the approach presented here,  entanglement can be distributed in a quantum network without the need for single-photon sources or detectors.
\end{abstract}

\maketitle

Fault-tolerant quantum computing will require redundancy to identify and correct  errors during a computation. In most architectures, the physical qubits will therefore vastly outnumber the logical qubits. The need to scale up existing  architectures has motivated a network approach  where remote qubits, grouped into  nodes, are connected by quantum-photonic interconnects~\cite{northup2014quantum,reiserer2015cavity,vandersypen2017interfacing,awschalom2021development, ruf2021quantum}. These  quantum networks naturally require entanglement distribution across nodes. Consequently, significant  effort has gone towards generating both heralded~\cite{duan2004scalable,hofmann2012heralded,bernien2013heralded, hensen2015loophole,delteil2016generation,narla2016robust, pompili2021realization}  and deterministic~\cite{cirac1997quantum,ritter2012elementary,kurpiers2018deterministic, axline2018demand, campagne2018deterministic, leung2019deterministic, magnard2020microwave,zhong2021deterministic}  qubit-photon entanglement. 

\begin{figure}
    \centering
    \includegraphics[width=\linewidth]{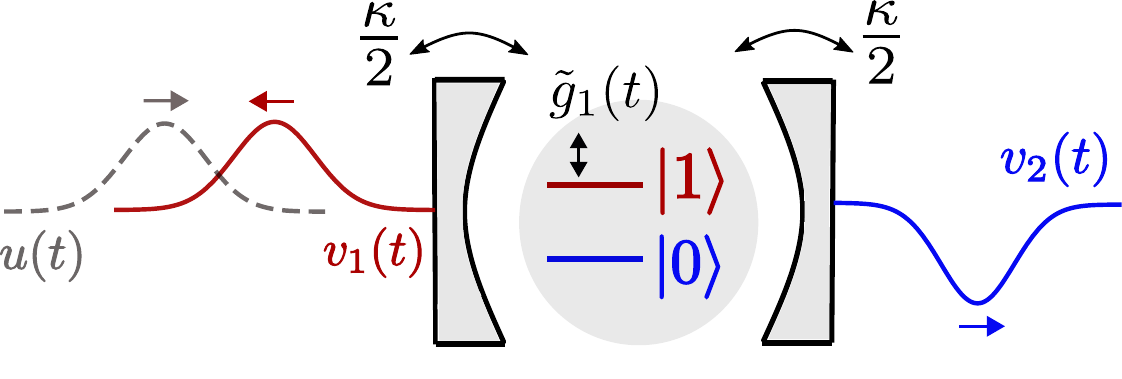}
    \caption{In the presence of a longitudinal cavity-qubit coupling modulated at the cavity frequency with envelope $\tilde{g}_1(t)$, an incoming coherent state with waveform $u(t)$ is  reflected (transmitted) into a coherent state with waveform $v_1(t)$ [$v_2(t)$] for control-qubit state $\ket{1}$ ($\ket{0}$). This effect requires symmetric decay rates $\kappa_1=\kappa_2=\kappa/2$ for cavity ports $i=1,2$.}
    \label{fig:setup}
\end{figure}

In this Letter, we present a photonic which-path entangler  that correlates the path of an incoming  multiphoton coherent-state wavepacket  with the state of a cavity-coupled control qubit (Fig.~\ref{fig:setup}).  The resulting   which-path degree of freedom, consisting of a coherent-state wavepacket traveling in one of two transmission lines, can be re-encoded in the photon-number parity of a continuous-variable degree-of-freedom, then used to generate entanglement with  one or more distant  qubits.  The entangler presented here therefore provides a natural interface between discrete- and  continuous-variable approaches to hybrid quantum computation~\cite{andersen2015hybrid, takeda2015entanglement, huang2019engineering, guccione2020connecting, djordjevic2022hybrid,darras2023quantum}. The qubit--which-path (QWP) state generated by the entangler also has greater potential sensitivity for phase measurements than either the comparable entangled coherent state (ECS)~\cite{sanders1992entangled,joo2011quantum} (consisting of a superposition of coherent states, one in each  interferometer arm)  or NOON state~\cite{bollinger1996optimal,lee2002quantum,giovannetti2011advances} (an analogous superposition of $N$-photon Fock states).  Quantum-enhanced interferometry has applications in, e.g., biological imaging~\cite{taylor2016quantum, moreau2019imaging,mukamel2020roadmap} and gravitational wave detection~\cite{schnabel2010quantum,aasi2013enhanced, tse2019quantum,yu2020quantum}.

A key requirement for the entangler is a modulated longitudinal (qubit-eigenstate preserving) cavity-qubit coupling.  Longitudinal coupling has attracted significant theoretical and experimental attention in recent years, as it is currently being realized and leveraged in a number of promising quantum-computing architectures~\cite{kerman2013quantum,billangeon2015circuit,didier2015fast,beaudoin2016coupling,richer2016circuit,richer2017inductively,harvey2018coupling,lambert2018amplified,bosco2022fully,bottcher2022parametric,michal2023tunable}. Though many current implementations of, e.g., cavity-coupled spin, charge, and superconducting qubits make use of the (transverse)  Rabi coupling, longitudinal cavity-qubit couplings are no less fundamental. They can be engineered for single-electron-spin qubits in  double quantum dots (DQDs), which can be coupled to cavity electric fields via magnetic-field gradients~\cite{beaudoin2016coupling}, as well as for hole-spin qubits in semiconductors~\cite{bosco2022fully, michal2023tunable, fang2023recent}, which interact with electric fields via their large intrinsic spin-orbit coupling.  Two-electron-spin singlet-triplet qubits in DQDs can be longitudinally coupled to electric fields by modulating a gate voltage controlling the strength of the exchange interaction~\cite{harvey2018coupling,bottcher2022parametric}. Longitudinal coupling can also be engineered in various superconducting-qubit architectures~\cite{kerman2013quantum, billangeon2015circuit, didier2015fast,richer2016circuit, richer2017inductively}. Moreover, even in systems where the dominant source of cavity-qubit coupling is intrinsically transverse,  an effectively longitudinal interaction can be engineered (in some rotating frame) by modulating the coupling strength at both the cavity and qubit frequencies~\cite{lambert2018amplified}, making the theory presented here widely applicable.

\textit{Model}---A longitudinal cavity-qubit interaction arises, e.g., from the DC Stark shift due to electric dipole coupling ($\propto Ey$ for a cavity electric field $E$ polarized along $\hat{y}$). Quantizing the cavity field (focusing on a single cavity mode of frequency $\omega_\mathrm{c}$ and annihilation operator $a$), and in adiabatic perturbation theory, we find an interaction proportional to the product of $E\propto i(a^\dagger-a)$ and the dipole matrix element $\bra{s(t)}y\ket{s(t)}$ taken with respect to the instantaneous qubit energy eigenstate $\ket{s(t)}$ ($s=0,1$) \cite{supplement}. This gives an effective Hamiltonian $H_\mathrm{eff}(t)=\sum_s ig_s(t)(a^\dagger-a)\ket{s}\bra{s}$, where $\ket{s}$ is a time-independent state in the adiabatic frame, and where $g_s(t)=g_s[\{x_j(t)\}]\propto\bra{s(t)}y\ket{s(t)}$ inherits time dependence from a collection of control parameters $\{x_j\}$ that determine $\ket{s(t)}$. For spin qubits, the spin-dependent electric dipole matrix elements [and consequently $g_s(t)$] can be modulated through external electric fields or gate voltages~\cite{beaudoin2016coupling,bosco2022fully, michal2023tunable}. An analogous mechanism exists for flux-tunable superconducting transmon qubits, in which the couplings $g_s(t)$ can instead be tuned by modulating a flux~\cite{didier2015fast}. In what follows, we assume a time-independent value $g_0(t)=\bar{g}_0$ and a sinusoidal modulation of $g_1(t)$ at the cavity frequency, $g_1(t)=\bar{g}_1+2|\tilde{g}_1(t)|\cos\left[\omega_{\mathrm{c}} t-\vartheta(t)\right]$, where $\tilde{g}_1(t)=e^{i\vartheta(t)}\lvert\tilde{g}_1(t)\rvert$ is a slowly varying envelope with $\tilde{g}_1(0)\simeq0$ and duration $\tau$. (See the Supplementary Material, Ref.~\onlinecite{supplement}, for a sufficient condition on the parameters $\{x_j\}$ in general, as well as specific conditions to achieve this coupling modulation for double-quantum-dot charge and spin qubits.) A polaron transformation can then be used to eliminate the term $\propto \bar{g}_0$ by incorporating a small shift $\sim \bar{g}_0^2/\omega_c$ in the qubit frequency $\omega_{\mathrm{q}}$. Going to an interaction picture with respect to the decoupled Hamiltonian $\omega_{\mathrm{c}}a^\dagger a +\omega_{\mathrm{q}}\sigma_z/2$ $(\sigma_z=\ket{0}\bra{0}-\ket{1}\bra{1})$, and within a rotating-wave approximation requiring that $\lvert \bar{g}_1\rvert, \lvert \tilde{g}_1(t)\rvert \ll \omega_{\mathrm{c}}$, the  cavity-qubit Hamiltonian is then given by \cite{supplement}
\begin{equation}
    H_0(t)=\frac{\xi(t)}{2}\sigma_z+i\ket{1}\bra{1}\left[\tilde{g}_1(t)a^\dagger-\mathrm{h.c.}\right],\label{hamiltonian}
\end{equation}
where we have introduced a stochastic noise parameter $\xi(t)$  leading to qubit dephasing. In general, the dipole approximation also produces a transverse Rabi term [$ig_\perp\sigma_x(a^\dagger-a)$], which, in the regime $|g_\perp|< \lvert\delta\rvert$ ($\delta=\omega_{\mathrm{q}}-\omega_{\mathrm
c}$), leads to a dispersive coupling $\chi \sigma_z a^\dagger a$, where $\chi=g_\perp^2/\delta$. Any effects due to transverse coupling can be suppressed by operating in the regime $\lvert g_\perp\rvert\ll \lvert \delta\rvert$.

The longitudinal interaction  $\propto \tilde{g}_1(t)$ displaces the cavity vacuum into a finite-amplitude coherent state for $s=1$. A similar effect is studied in Ref.~\onlinecite{didier2015fast} to design a fast quantum non-demolition (QND) qubit readout. Relative to Ref.~\onlinecite{didier2015fast}, we additionally consider driving of the cavity by an input field. In particular, we assume that the cavity is  coupled to  external  transmission lines at  input ($i=1$) and output ($i=2$) ports (Fig.~\ref{fig:setup}). An input spatiotemporal mode (wavepacket) with normalized waveform $u(t)$   [$\int dt\lvert u(t)\rvert^2=1$]
can be represented by the mode operator $b_u=\int dt\:u^*(t)r_{\mathrm{in},1}(t)$~\cite{kiilerich2019input,kiilerich2020quantum}, where $r_{\mathrm{in},i}(t)$  satisfies the input-ouput relation $r_{\mathrm{out},i}(t)=r_{\mathrm{in},i}(t)+\sqrt{\kappa_i}a(t)$~\cite{gardiner1985input}. Here, $r_{\mathrm{out},i}(t)$ is the output field, and $\kappa_i$ is the rate of decay from cavity port $i$. We assume that the quantum state of the incoming wavepacket is a coherent state with initial amplitude $\braket{b_u}=\alpha_0$, giving $\braket{r_{\mathrm{in},1}}_t=u(t)\alpha_0$. Where it appears, the notation $\braket{\mathcal{O}}_t$ indicates an average with respect to the initial state $\rho(0)$: $\braket{\mathcal{O}}_t=\mathrm{Tr}\{\mathcal{O}(t)\rho(0)\}$ for operator $\mathcal{O}$.  The reflected  and transmitted  waveforms  are given in the frequency domain by $v_1(\omega)=R(\omega)u(\omega)$ and $v_2(\omega)=T(\omega)u(\omega)$, respectively, where $R(\omega)=\braket{r_{\mathrm{out},1}}_\omega/\braket{r_{\mathrm{in},1}}_\omega$ and $T(\omega)=\braket{r_{\mathrm{out},2}}_\omega/\braket{r_{\mathrm{in},1}}_\omega=\sqrt{\kappa_2}\braket{a}_\omega/\alpha_0u(\omega)$ are the reflection and transmission coefficients with  $\braket{\mathcal{O}}_\omega=\int dt\:e^{i\omega t}\braket{\mathcal{O}}_t$.

To derive the transmission $T(\omega)$ conditioned on the qubit state $\ket{s}$, we now find $\braket{a}_\omega$ from the quantum Langevin equation for $\braket{a}_t$,
\begin{equation}
\braket{\dot{a}}_t=-\frac{\kappa}{2}\braket{a}_t+\tilde{g}_1(t)s-\sqrt{\kappa_1}\alpha_0u(t).    \label{langevin}
\end{equation}
The displacement of the cavity vacuum due to the interaction $\propto \tilde{g}_1(t)$ can therefore be canceled exactly, conditioned on the qubit being in state $\ket{1}$, by ensuring that $\sqrt{\kappa_1}\alpha_0u(t)=\Tilde{g}_1(t)$. Destructive interference then precludes a transfer of photons to the output transmission line, leading to perfect reflection of the input field. Evidence of such destructive interference was recently observed experimentally in Ref.~\onlinecite{corrigan2022longitudinal}, where a modulated longitudinal coupling and a cavity drive were both generated with a common voltage source (acting as a common phase reference). Because the input state is a coherent state [and coherent states are eigenstates of $r_{\mathrm{in},1}(t)$],  there are no quantum fluctuations about the average dynamics  $\braket{r_{\mathrm{in},1}}_t=\alpha_0 u(t)$. For a non-ideal input, however, fluctuations about the average ($\alpha_0\rightarrow\alpha_0+\delta\alpha$) lead to imperfect cancellation  for $s=1$.

For a cavity that is initially empty, we have $\braket{a}_0=0$. Integrating  the quantum Langevin equation [Eq.~\eqref{langevin}] with this initial condition gives
\begin{equation}
    \braket{a}_\omega=\chi_{\mathrm{c}}(\omega)[\tilde{g}_1(\omega)s-\sqrt{\kappa_1}\alpha_0u(\omega)],\label{cavityfield}
\end{equation}
where $\chi_{\mathrm{c}}(\omega)=(\kappa/2-i\omega)^{-1}$. For $\sqrt{\kappa_1}\alpha_0u(\omega)=\tilde{g}_1(\omega)$, the transmission  can then be written as
\begin{equation}
    T(\omega)=(1-s)\frac{\sqrt{\kappa_1\kappa_2}}{i\omega-\kappa/2}.\label{Ts}
\end{equation}
The input pulse $u(\omega)$ has support for $\omega\lesssim 1/\tau$ localized about the cavity frequency (corresponding to $\omega=0$ in the rotating frame). Near-perfect transmission  can then be  achieved for $s=0$ and $\kappa_1=\kappa_2=\kappa/2$ by operating in the regime  of large $\kappa\tau$, where $\chi_{\mathrm{c}}(\omega)$ is much broader in frequency than $u(\omega)$. Finite-bandwidth effects for a Gaussian input waveform $u(t)$ may be neglected provided~\cite{supplement} $N=\lvert \alpha_0\rvert^2\ll (\kappa\tau)^4$. Given $T(\omega)$ for a fixed value of $s$,  $R(\omega)$ is related to $T(\omega)$ through the input-output relation, $\sqrt{\kappa_2}[R(\omega)-1]=\sqrt{\kappa_1}T(\omega)$. 

An alternative way to condition the cavity transmission on the state of a qubit would be to engineer a qubit-state-dependent shift of the cavity frequency  using dispersive coupling $\chi\sigma_za^\dagger a$, where $2\lvert \chi\rvert\gg\kappa$. A narrow-band input tone at frequency $\chi$ would then be transmitted conditioned on state $\ket{0}$ and reflected for state $\ket{1}$. However, in the dispersive regime ($\epsilon=\lvert g_\perp/\delta\rvert<1$), this necessarily requires (very) strong coupling $\lvert g_\perp\rvert\gg \kappa/\epsilon$. The entangler presented here, by contrast, can be operated even if $\lvert\Tilde{g}_1\rvert\lesssim \kappa$. Dipole-induced transparency \cite{waks2006dipole} and reflection~\cite{auffevesgarnier2007giant} also result in perfect steady-state transmission or reflection of a weak input pulse conditional on the presence of a resonant, transversally coupled dipole. These effects are not, however, QND in the state of the decoupled dipole and furthermore require that the cavity be driven with an average of $N_\mathrm{cav}\lesssim 1$ intracavity photon~\cite{englund2007controlling}. For the entangler presented here, by contrast, the transmission vs reflection of a transient pulse is QND; it is conditioned on the decoupled qubit state $\ket{s}$. Moreover, the entangler works in the regime $N_\mathrm{cav}\sim N/(\kappa\tau)>1$, provided the finite-bandwidth condition is satisfied [$N\ll (\kappa\tau)^4$ for a Gaussian waveform]. 

The qubit-state-conditioned transmission [Eq.~\eqref{Ts}] can be used to generate entangled states. To describe the states associated with the reflected and transmitted fields, we use the virtual-cavity formalism  of Refs.~\cite{kiilerich2019input, kiilerich2020quantum}  to recast the input, reflected, and transmitted wavepackets as the fields emitted from---or absorbed into---fictitious (virtual) single-sided cavities coupled to the transmission lines  via time-dependent couplings. This formalism allows for an efficient description of the scattering of an input pulse into pre-specified spatiotemporal modes, which can be modeled as the modes of virtual cavities. Accounting for the input pulse, cavity field, reflected pulse, and transmitted pulse, the evolution of the cavity and transmission lines is then fully described by an effective four-mode model. 
 The quantum master equation governing this evolution is~\cite{kiilerich2019input,kiilerich2020quantum}  
\begin{equation}
    \dot{\rho}_\xi=-i[H_0(t)+V(t),\rho_\xi]+ \sum_{i=1,2}\mathcal{D}[L_i]\rho_\xi,\label{master-eq}
\end{equation}
where $\rho_\xi$ is the density matrix conditioned on a realization of the noise $\xi(t)$, and where
\begin{align}
    V(t)&=\frac{i}{2}[\sqrt{\kappa_1}\lambda_u^*(t) a_u^\dagger a+\sum_{i=1,2}\sqrt{\kappa_i}\lambda_{v_i}(t)a^\dagger a_{v_i}\nonumber\\
    &+\lambda_u^*(t)\lambda_{v_1}(t)a_u^\dagger a_{v_1}-\mathrm{h.c.}].\label{coupling}
\end{align}
In Eq.~\eqref{master-eq}, $\mathcal{D}[L]\rho=L\rho L^\dagger-\frac{1}{2}\{L^\dagger L,\rho\}$ is a damping superoperator.  The operators $L_1=\sum_{\mu=u,v_1}\lambda_\mu(t)a_\mu+\sqrt{\kappa_1}a$ and $L_2=\lambda_{v_2}(t)a_{v_2}+\sqrt{\kappa_2}a$ model decay from the virtual cavity modes $a_\mu$, as well as decay out of the  cavity mode $a$ with  rate $\kappa=\kappa_1+\kappa_2$. The couplings to the virtual cavities are given by $\lambda_u(t)=u(t)/(\int_t^\infty dt'\lvert u(t')\rvert^2)^{1/2}$ and 
$\lambda_{v_i}(t)=(-1)^iv_i(t)/(\int_0^tdt'\lvert v_i(t')\rvert^2)^{1/2}$~\cite{kiilerich2019input,kiilerich2020quantum,nurdin2016perfect}. The $i$-dependent sign in $\lambda_{v_i}$ reflects the fact that in our model, a transmitted pulse undergoes a $\pi$ phase shift [cf Eq.~\eqref{Ts}]. The couplings $\lambda_u(t)$ and $\lambda_{v_i}(t)$ have singularities at $t\rightarrow\infty$ and $t=0$, respectively, and, for real cavities, both $\lambda_u(t)$ and $\lambda_{v_i}(t)$ would have to be truncated to finite values to realize absorption (emission) into (out of) a chosen spatiotemporal mode~\cite{nurdin2016perfect}. In the virtual-cavity formalism, however, these unphysical couplings are abstractions used to calculate the dynamics into/out of a chosen mode. No additional (real) cavities or time-dependent couplings are required to realize the entangler.

We assume a fast $\pi/2$ pulse can be used to prepare the qubit in the $(t=0)$ initial state $\ket{+}=(\ket{1}+\ket{0})/\sqrt{2}$, with the cavity in the vacuum state ($\left<a\right>_0=0$). A coherent-state wavepacket then evolves from the input mode $a_u$ to the two output modes $a_{v_i}$ ($i=1,2$). For times $t\gg \tau$  exceeding the duration  of the input pulse, the quantum states associated with the reflected and transmitted waveforms $v_i(t)$, conditioned on $s$, will have been fully transferred into their respective fictitious cavities: $\alpha_{is}=\mathrm{lim}_{t\rightarrow\infty}\braket{a_{v_{i}}}_t=(-1)^{i-1}\alpha_0\int \frac{d\omega}{2\pi} \lvert v_{i}(\omega)\rvert^2$. The joint state of the qubit and transmission lines, found from a direct integration of Eq.~\eqref{master-eq}, is then $\rho(t)=\llangle \rho_\xi(t)\rrangle$, where $\llangle\rrangle$ denotes an average over realizations of $\xi(t)$, and where $\rho_\xi(t)=\ket{\Psi_{\xi(t)}}\bra{\Psi_{\xi(t)}}$ with 
\begin{equation}
    \ket{\Psi_{\xi(t)}}=\frac{1}{\sqrt{2}}\left(e^{\frac{i}{2}\theta_\xi(t)}\ket{1,\psi_{\mathrm{1}}}+e^{-\frac{i}{2}\theta_\xi(t)}\ket{0,\psi_{\mathrm{0}}}\right).\label{which-path}
\end{equation}
Here, $\theta_\xi(t)=\int_0^t dt'\xi(t')$ is a random phase, $\ket{\psi_s}=\prod_{i=1,2}D_{i}(\alpha_{is})\ket{\text{vac}}$ is the state of the transmission lines conditioned on $s$, $\ket{\text{vac}}$ is the  vacuum, and $D_i(\alpha)=\mathrm{exp}\{\alpha a_{v_i}^\dagger-\mathrm{h.c.}\}$ is a displacement operator.  For $\kappa_1=\kappa_2=\kappa/2$ and up to corrections in $N/(\kappa\tau)^4\ll 1$, only one of $\alpha_{is}$ is nonzero for each value of $s$: For $s=1$, $\alpha_{11}=\alpha_0$ and $\alpha_{21}=0$, while for $s=0$, $\alpha_{10}=0$ and $\alpha_{20}=-\alpha_0$.  Equation~\eqref{which-path} therefore describes a photonic which-path qubit entangled with the control qubit (a QWP state). Under the same finite-bandwidth conditions, imperfections in the input source such that  $\alpha_0\rightarrow \alpha_0+\delta\alpha$ will lead instead to $\alpha_{11}=\alpha_0$, $\alpha_{21}=-\delta\alpha$, $\alpha_{10}=0$, and $\alpha_{20}=-(\alpha_0+\delta\alpha)$. This follows from integrating the Langevin equation [Eq.~\eqref{langevin}] with $\alpha_0\rightarrow \alpha_0+\delta\alpha$ and solving for the reflected and transmitted fields. If we take $\delta\alpha$ to be a complex-valued, zero-mean Gaussian random variable, then the fidelity of the ideal QWP state [Eq.~\eqref{which-path} with $\xi=0$] with respect to the mixed state obtained by averaging over $\delta\alpha$ is $e^{-\braket{\lvert \delta\alpha\rvert^2}_{\delta\alpha}}$, where here, $\braket{}_{\delta\alpha}$ describes an average over $\delta\alpha$. High-fidelity QWP states therefore require a stable coherent-state source with a low absolute noise level, below one photon per pulse ($\braket{\lvert\delta\alpha\rvert^2}_{\delta\alpha}\ll1$).

\begin{figure}
    \centering
    \includegraphics[width=\linewidth]{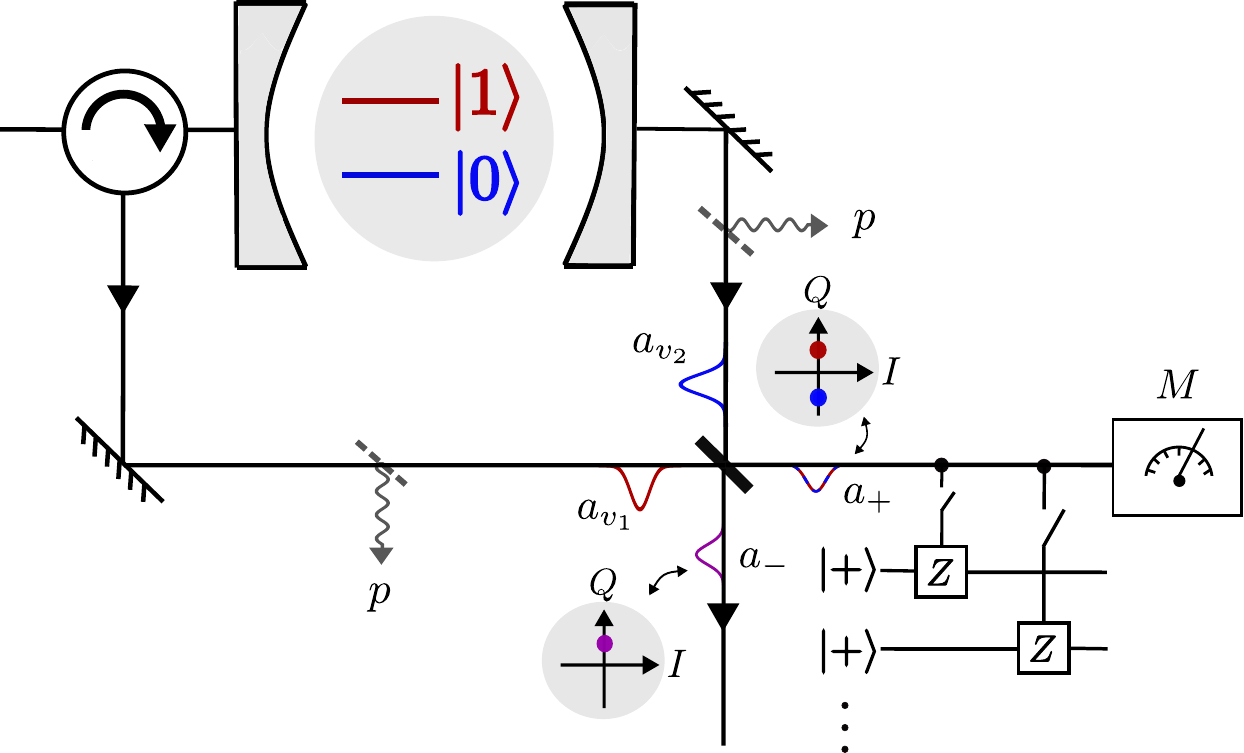}
    \caption{ An interferometry setup can be used to entangle the which-path degree of freedom with a target qubit initialized in $\ket{+}$ via a conditional phase shift: $Z\ket{+}=\ket{-}$. This can be accomplished by re-encoding the which-path degree-of-freedom in the photon-number parity of a cat state (composed of a superposition of two distinct coherent states occupying the spatiotemporal mode annihilated by $a_+$) propagating to the right of the 50:50 beamsplitter. The symbol labeled $M$ represents a measurement of the quadrature $Q$, which can be used to prepare an $m$-qubit GHZ state (or Bell state for $m=2$) involving the control qubit and $(m-1)$  target qubits. Photon loss occurring with probability $p$ is modeled with fictitious beamsplitters having reflectivity $p$.}
    \label{fig:measurement}
\end{figure}

\textit{Entanglement distribution}---The which-path degree-of-freedom  can  be entangled with a second (``target'') qubit for long-range entanglement distribution. Crucially, this also provides a direct avenue for quantifying entanglement in QWP states through measurements of stationary qubits only. The setup is illustrated schematically in Fig.~\ref{fig:measurement}.  By interfering the reflected and transmitted fields at a 50:50 beamsplitter, the output modes can be mapped to new modes $a_\pm$ such that $\braket{a_{\pm ,s}}=(\alpha_{1s}\pm \alpha_{2s})/\sqrt{2}$.  This gives $\braket{a_{-,s}}= \alpha$ independent of $s$, where $\alpha=\alpha_0/\sqrt{2}$.  The $a_+$ mode, by contrast, has $s$-dependence given by $\braket{a_{+,s}}= (2s-1)\alpha$. The beamsplitter consequently re-encodes the which-path degree-of-freedom  in the phase of the coherent-state amplitude: $\braket{a_{+,1}}=+\alpha$  and $\braket{a_{+,0}}=-\alpha$.  

For clarity, we now set $\xi(t)=0$ for the purpose of explaining the protocol. The effects of dephasing [$\xi(t)\neq 0$] are included in the relevant result, Eq.~\eqref{concurrence}, below. Since the $a_-$ mode does not share entanglement with either the qubit or $a_+$ mode, it can be measured (traced over) without disturbing the state of the qubit and $a_+$ mode. The post-measurement state is then $(1/2)\sum_{\lambda=\pm}\sqrt{\mathcal{N}_\lambda}\ket{\lambda,C_\lambda}$, where  $\ket{C_\pm}=(\ket{+\alpha}\pm \ket{-\alpha})/\sqrt{\mathcal{N}_\pm}$ are cat states (consisting of only even or odd photon-number states) and $\mathcal{N}_\pm$ are normalization factors.  With the target qubit initialized in $\ket{+}$, the control and target qubits can  be entangled through a  phase flip on the target, $\ket{+}\rightarrow\ket{-}$, conditioned on an odd photon-number parity~\cite{besse2018single,kono2018quantum,hacker2019deterministic,besse2020parity,supplement}. Following such a phase flip, the state of the qubits and electric field is  $(1/2)\sum_{\lambda=\pm}\sqrt{\mathcal{N}_\lambda}\ket{\lambda,\lambda,C_\lambda}$. In the limit  $\braket{\alpha\vert -\alpha}\rightarrow 0$, a final quadrature measurement of the electric field can then be used to project the  qubits into the Bell state $(\ket{++}\pm\ket{--})/\sqrt{2}$, conditioned on outcome $\ket{\pm \alpha}$. Multi-qubit Greenberger-Horne-Zeilinger (GHZ) states $\ket{+,+,...,+}\pm \ket{-,-,...,-}$ can also be generated by allowing the field to interact sequentially with a series of potentially distant qubits. In contrast to the well-established single-photon pitch-and-catch approach to long-range entanglement distribution, involving the emission and destructive reabsorption of a single photon~\cite{cirac1997quantum,ritter2012elementary,kurpiers2018deterministic, axline2018demand, campagne2018deterministic, leung2019deterministic, magnard2020microwave}, the approach presented here is QND in the photon number and therefore provides a clear path towards the generation of multipartite entangled states.
 
We can quantify the effects of photon loss on the amount of distributed entanglement by inserting a
fictitious beamsplitter, described by the unitary $B_{c,c'}(\varphi)=e^{i\varphi(c^\dagger c'+\mathrm{h.c.})}$ \cite{joo2011quantum}, into each arm of the interferometer (Fig.~\ref{fig:measurement}). With $c =a_{v_i}$, this describes scattering from $a_{v_i}$ into loss mode $a_{v_i}'$ (in the environment) with probability $p=\mathrm{cos}^2\varphi$. Tracing over the loss modes $a_{v_{1,2}}'$ then yields a reduced density matrix for the state of the qubit and modes $a_{v_{1,2}}$. In the presence of such loss, and for finite coherent-state overlap $\braket{\alpha\vert-\alpha}\neq 0$, the post-measurement state of the qubits is not an ideal Bell state, but is instead mixed: For a measurement of the electric field along the quadrature $Q$ of coherent-state displacement~\cite{dragan2001homodyne}, the post-measurement state of the qubits (conditioned on an inferred displacement along $\pm Q$) is an X-state~\cite{yu2005evolution} whose concurrence $C$~\cite{hill1997entanglement, wootters1998entanglement, wootters2001entanglement} can easily be computed \cite{supplement}:
\begin{equation}
    C(t)=\mathrm{max}\{0,\mathrm{erf}(\sqrt{N_{\eta}})e^{-N_{p}-\chi_\xi(t)}-\mathrm{erfc}(\sqrt{N_{\eta}})\},\label{concurrence}
\end{equation}
 where $N_{p}=pN=p \lvert \alpha_0\rvert^2$ and $N_{\eta}=\eta (1-p)N$ are controlled by the average number of photons lost and detected, respectively. Here, $\eta\in (0,1]$ is the detector efficiency, and
 \begin{equation}
     \chi_\xi(t)=\int \frac{d\omega}{2\pi }\frac{4\mathrm{sin}^2\left(\tfrac{\omega t}{2}\right)}{\omega^2} S(\omega)
 \end{equation}
 results from an average $\llangle \rrangle$ over realizations of the noise $\xi(t)$, here taken to be stationary, zero-mean Gaussian noise with spectral density $S(\omega)=\int dt\:e^{-i\omega t}\llangle \xi(t)\xi(0)\rrangle$. The concurrence [Eq.~\eqref{concurrence}] quantifies the amount of entanglement that can be distributed to a second qubit in the presence of (symmetric) photon loss in the interferometer, qubit dephasing, and imperfect assignment fidelity at the final measurement of the electric field. In particular, the expression for $C(t)$ indicates that for fixed values of $p$ and $\eta$, there is an optimal $N$ that maximizes the entanglement \cite{supplement}. In the presence of asymmetric losses with probabilities $p_1$ and $p_2$, Eq.~\eqref{concurrence} with $p=\mathrm{max}\{p_1,p_2\}$ provides a lower bound on the achievable concurrence.

 The requirement $\sqrt{\kappa/2}\alpha_0u(t)=\Tilde{g}_1(t)$ limits the average number of photons in the input coherent state. For a Gaussian $u(t)$, $N=\lvert\alpha_0\rvert^2= 2\sqrt{\pi}(\Tilde{g}_1^{\mathrm{max}})^2\tau/\kappa$, where $\Tilde{g}_1^{\mathrm{max}}=\mathrm{max}_t\lvert\Tilde{g}_1(t)\rvert$. Together with the bandwidth requirement $N\ll (\kappa\tau)^4$, this implies that $N\ll N_{\mathrm{max}}\equiv (\Tilde{g}_1^{\mathrm{max}}\tau)^{8/5}$. A larger $\tau$ therefore increases $N_{\mathrm{max}}$. However, the pulse duration is also subject to the requirement $\tau<T_2^*$, where $T_2^*$ is the dephasing time of the qubit [defined by $\chi_\xi(T_2^*)=1$]. For example, if  $g_1^\mathrm{max}/2\pi =1\,\mathrm{MHz}$ and $\tau=1\,\mu\mathrm{s}$, then $N_\mathrm{max}\simeq 19$. For $\tilde{g}_1/\kappa=1/8$ (weak coupling), we then have $N\simeq 3$, close to the value that maximizes the two-qubit concurrence ($C\simeq 0.95$) for $p=0.01$ \cite{supplement}. This scenario may be realistic for, e.g., an electron-spin qubit in a silicon DQD with a magnetic field gradient. The longitudinal coupling for this case can be comparable to the transverse coupling $\sim 1-10\,\mathrm{MHz}$ \cite{beaudoin2016coupling}. Dephasing times for electron-spin qubits in $^{28}$Si quantum dots reach $T_2^*\sim 100\,\mu\mathrm{s}\gg \tau$~\cite{stano2022review}. The same values of $N$ and $N_\mathrm{max}$ could also be realized for flux-tunable transmons, with a longitudinal coupling $\sim 10\,\mathrm{MHz}$ \cite{didier2015fast} and pulse duration $\tau\sim 100\,\mathrm{ns}\ll T_2^*$ (for transmons, coherence times reach $T_2^*\simeq 100\,\mu\mathrm{s}$ \cite{kjaergaard2020superconducting}).  
 
 \textit{Precision metrology}---The entangler described above can also be used to perform quantum-enhanced precision measurements of a phase $\phi$ acquired by the field reflected from the cavity as it propagates along arm 1 of an interferometer (Fig.~\ref{fig:measurement}). The fundamental precision bound for estimation of $\phi$ (the  quantum Cram\'er-Rao bound~\cite{braunstein1994statistical,braunstein1996generalized}) is better for QWP states than for either NOON states (superpositions of $N$-photon Fock states, one in each interferometer arm) or entangled coherent states \cite{sanders1992entangled}  (similarly, superpositions of coherent states) having the same average number $N$ of photons~\cite{inprep}.

\textit{Outlook}---The entangler presented here could also be used to perform measurements of the phase acquired by the control qubit. Specifically, a modulated longitudinal coupling, followed by a rapid reset~\cite{reed2010fast,johnson2012heralded,geerlings2013demonstrating, kobayashi2023feedback} $\ket{0}\rightarrow\ket{1}$, can be used to map the relative phase $\ket{0}+e^{i\theta}\ket{1}$ of the initial qubit state onto the state $\ket{-\alpha}+e^{i\theta}\ket{\alpha}$ of the $a_+$ mode. A projective measurement of  $\ket{C_\pm}$ then yields a single bit of information about $\theta$ (the maximum achievable for a single-shot qubit readout). This may be useful in situations where $\theta$ encodes information about dynamics induced by a classical or quantum environment~\cite{mcintyre2022non,mutter2022fingerprints}.

\begin{acknowledgments}
\emph{Acknowledgments}---We thank A.~Blais and K.~Wang for useful discussions. We also  acknowledge funding from the Natural Sciences and Engineering Research Council (NSERC) and from the Fonds de recherche--Nature et technologies (FRQNT).
\end{acknowledgments}

\providecommand{\noopsort}[1]{}\providecommand{\singleletter}[1]{#1}%

\clearpage
\onecolumngrid

\renewcommand{\thesection}{\Roman{section}}

\renewcommand{\theequation}{S\arabic{equation}}
\renewcommand{\thefigure}{S\arabic{figure}}
\renewcommand\bibnumfmt[1]{[S#1]}
\renewcommand{\citenumfont}{S}

\setcounter{secnumdepth}{3}
\setcounter{equation}{0}
\renewcommand{\thesection}{S\Roman{section}}

\begin{center}
{\large \textbf{Supplementary Material for `A photonic which-path entangler based on  longitudinal cavity-qubit coupling'}}\\\smallskip
Z. M. McIntyre and W. A. Coish\\
Department of Physics, McGill University, 3600 rue University, Montreal, QC, H3A 2T8, Canada
\end{center} 

\section{Cavity-qubit Hamiltonian}
\label{sec:Hamiltonian}

In general, the wavefunctions defining the ground and excited states of a qubit can be tuned through one or more parameters. For spin qubits in gate-defined double quantum dots, these could be gate voltages; for superconducting qubits, they could be fluxes. When these parameters are modulated in time, the Hamiltonian $H_{\mathrm{q}}$ of the qubit acquires a time dependence. The full cavity-qubit Hamiltonian can then be written as 
\begin{equation}
    H(t)=H_\mathrm{q}(t)+\omega_{\mathrm{c}}a^\dagger a+H_{\mathrm{int}},\label{S:hamiltonian}
\end{equation}
where $H_{\mathrm{q}}(t)\ket{s(t)}=\epsilon_s(t)\ket{s(t)}$ is the Hamiltonian whose low-energy instantaneous eigenstates (labelled by $s=0,1$) are used to encode the qubit, and where $a$ annihilates an excitation in the cavity mode (whose frequency is denoted $\omega_{\mathrm{c}}$). In many architectures (involving e.g. atoms, excitons, or spins together with spin-orbit coupling), the cavity-qubit interaction $H_{\mathrm{int}}$ has its origins in the electric-dipole interaction $H_{\mathrm{int}}=e \bm{E}\cdot \bm{r}$, where $e>0$ is the magnitude of the electron charge, $\bm{E}$ is the electric field of the cavity at the location of the dipole (qubit), and $\bm{r}$ is the position operator of an electron.

We transform $H(t)$ [Eq.~\eqref{S:hamiltonian}] via the unitary transformation $U(t)=\sum_{s}\ket{s}\bra{s(t)}$ to the instantaneous adiabatic eigenbasis, in which 
\begin{equation}
    \tilde{H}(t)=U(t)H(t)U^\dagger(t)-iU(t)\dot{U}^\dagger(t).
\end{equation}
So far, this transformation is exact. Treating the evolution of the system within an adiabatic approximation,
\begin{equation}
    \Tilde{H}(t)\simeq \tilde{H}_{\mathrm{adiabatic}}(t)=U(t)H(t)U^\dagger(t),
\end{equation}
requires that the usual adiabaticity condition be satisfied, 
\begin{equation}
       \frac{\lvert\braket{ \bar{s}(t)\vert \partial_t\rvert s(t)}\rvert}{\lvert \epsilon_0(t)-\epsilon_1(t)\rvert}\ll 1,
\end{equation}
where $s\in\left\{0,1\right\}$, $\bar{1}=0$, and $\bar{0}=1$. A similar condition must also be obeyed by proximal excited states. In particular, since we consider modulation at the cavity frequency $\omega_c$, there should be no coupled excited states at (or near) the cavity resonance. The cavity electric field is quantized $\bm{E}=i\bm{E_0}(a^\dagger-a)$, giving
\begin{equation}
    \tilde{H}_{\mathrm{adiabatic}}(t)=\epsilon_0(t)\ket{0}\bra{0}+\epsilon_1(t)\ket{1}\bra{1}+\omega_{\mathrm{c}}a^\dagger a + i\sum_{s=0,1}g_s(t)\ket{s}\bra{s}(a^\dagger-a)+i\sum_{s}g_\perp(t)\ket{s}\bra{\bar{s}}(a^\dagger-a),\label{S:H-adiabatic-1}
\end{equation}
where the longitudinal coupling $g_s(t)=e\bm{E_0}\cdot\braket{s(t)\lvert \bm{r}\lvert s(t)}$ depends on the $s$-dependent electric dipole moment $-e\braket{s(t)\lvert \bm{r}\lvert s(t)}$, and where the transverse coupling $g_\perp(t)=e\bm{E_0}\cdot\braket{s(t)\lvert  \bm{r}\rvert \bar{s}(t)}$ depends on the transition dipole moment $-e\braket{s(t)\lvert \bm{r}\lvert \bar{s}(t)}$. When the transition dipole is weak, or for a qubit strongly detuned from the cavity, the transverse term may be strongly suppressed, yielding an interaction that is predominantly longitudinal.

We assume that $g_0(t)$ and $g_1(t)$ can both be tuned by varying parameters $x_j(t)$. (These could be local potentials or electric fields for atomic, exciton, or spin qubits, or they could be fluxes for flux-biased superconducting qubits.) To linear order, 
\begin{equation}
    g_s(t)=\bar{g}_s+\delta g_s(t)\simeq \bar{g}_s+\sum_{j} \frac{\partial g_s}{\partial x_j}\delta x_j(t),\label{S:gs(t)}
\end{equation}
where $\delta g_s(t)$ describes a time-dependent modulation of $g_s(t)=\bar{g}_s+\delta g_s(t)$ [and similarly for $\delta x_i(t)$].

When one of $g_s(t)$ (for $s=0,1$) can be modulated independently of the other with a single $x_j(t)$, the required time-dependent modulation can be realized directly. More generally, it may be necessary to consider control through a minimum of two parameters $x_j(t)$ ($j=0,1$). Realizing the time-dependence of $g_s(t)$ required for the entangler then calls for a solution to the system of equations 
\begin{equation}
    \begin{pmatrix}
        \delta x_0\\
        \delta x_1
    \end{pmatrix}=\bm{J}^{-1}\begin{pmatrix}
        \delta g_0(t)\\
        \delta g_1(t)
    \end{pmatrix}=\bm{J}^{-1}\begin{pmatrix}
        0\\ 2\lvert\Tilde{g}_1(t)\rvert\mathrm{cos}[\omega_{\mathrm{c}}t-\vartheta(t)]
    \end{pmatrix},\label{system-of-eq}
\end{equation}
where here, $\bm{J}$ is a $2\times 2$ matrix with elements $J_{sj}=\partial g_s/\partial x_j$, and where $\Tilde{g}_1(t)=e^{i\vartheta(t)}\lvert\Tilde{g}_1(t)\rvert$ varies slowly relative to the timescale $\omega_{\mathrm{c}}^{-1}$. A sufficient condition for Eq.~\eqref{system-of-eq} to have a nontrivial solution is that $\bm{J}^{-1}$ be defined, i.e.~that the determinant of $\bm{J}$ be nonvanishing.

We assume that the modulations of $\delta x_j(t)$ prescribed by Eq.~\eqref{system-of-eq} have a negligible impact on $\epsilon_s(t)\simeq \epsilon_s$, so that, neglecting the transverse coupling term ($\propto g_\perp$) and up to a constant shift in energy,
\begin{equation}
    \Tilde{H}_{\mathrm{adiabatic}}(t)\simeq \frac{1}{2}\omega_{\mathrm{q}}'\sigma_z+\omega_{\mathrm{c}}a^\dagger a + ig_0\ket{0}\bra{0}(a^\dagger-a)+ig_1(t)\ket{1}\bra{1}(a^\dagger-a),\label{S:hamiltonian-adiabatic}
\end{equation}
where here, $\sigma_z=\ket{0}\bra{0}-\ket{1}\bra{1}$ and $\omega_{\mathrm{q}}'=\epsilon_0-\epsilon_1$. 

We now perform a polaron transformation $\dbtilde{H}(t)=e^S\tilde{H}_{\mathrm{adiabatic}}(t)e^{-S}$ on Eq.~\eqref{S:hamiltonian-adiabatic}, where
\begin{equation}
    S= i\frac{g_0}{\omega_{\mathrm{c}}}\ket{0}\bra{0}(a^\dagger-a).
\end{equation}
Up to a constant shift in energy, the result of this transformation is 
\begin{equation}
    \dbtilde{H}(t)= \frac{1}{2}\omega_{\mathrm{q}}\sigma_z+\omega_{\mathrm{c}}a^\dagger a+ig_1(t)\ket{1}\bra{1}(a^\dagger-a),\label{hamiltonian-polaron}
\end{equation}
where here, $\omega_{\mathrm{q}}=\omega_{\mathrm{q}}'-g_0^2/2\omega_{\mathrm{c}}$ is the qubit frequency accounting for the polaron shift. To model the effects of dephasing, we assume that the qubit splitting $\omega_{\mathrm{q}}$ in Eq.~\eqref{hamiltonian-polaron} is subject to classical fluctuations described by a stochastic noise process $\xi(t)$: $\omega_{\mathrm{q}}\rightarrow \omega_{\mathrm{q}}+\xi(t)$. Transforming Eq.~\eqref{hamiltonian-polaron} to an interaction picture with respect to $\omega_{\mathrm{q}}\sigma_z/2+\omega_{\mathrm{c}}a^\dagger a$ and performing a rotating-wave approximation then recovers Eq.~\eqref{hamiltonian} of the main text.

\subsection{Physical examples}
\begin{figure}
    \centering
    \includegraphics[width=\textwidth]{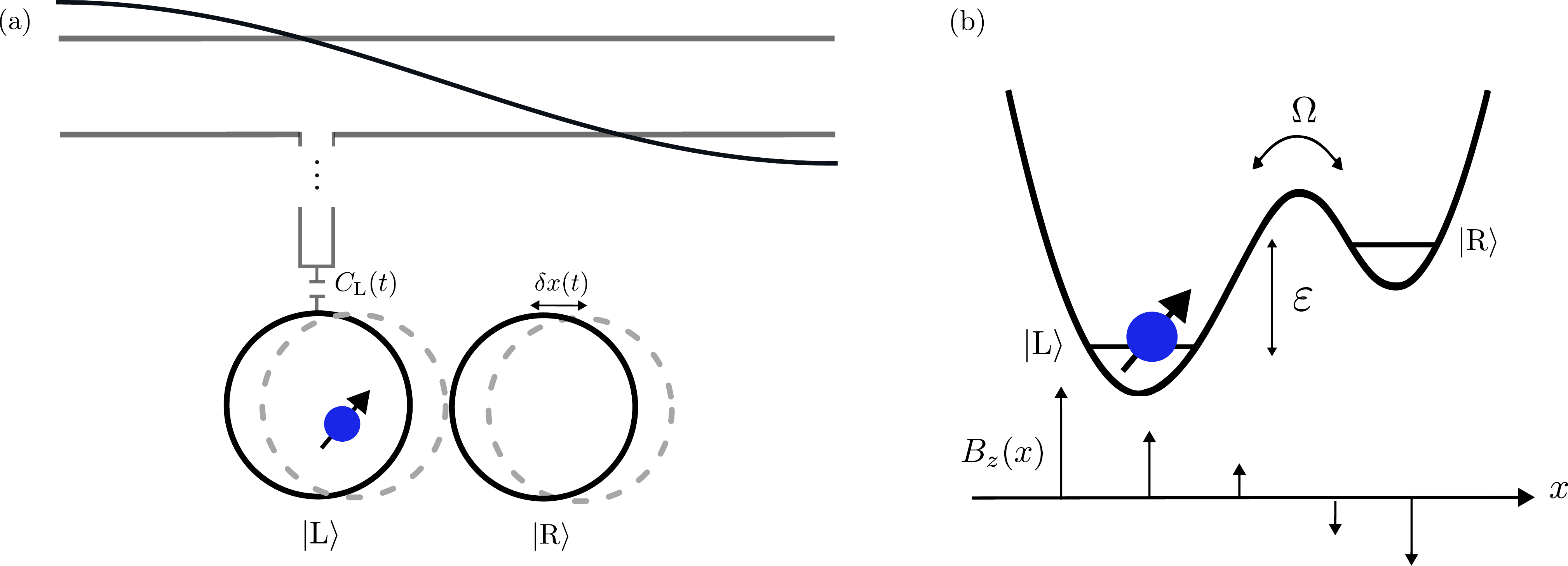}
    \caption{(a) An electron in a double quantum dot can be coupled capacitively to a microwave resonator via a metallic finger (vertical gray rectangle) located near one of the dots. We assume that the dot-to-finger lever arm $\alpha_{\mathrm{L}}\propto C_{\mathrm{L}}$ of the left dot can be modulated by shifting the position $x(t)$ of the orbitals $\psi_l(\boldsymbol{r},t)$ ($l=\mathrm{L,R}$) of the individual dots through a gate voltage modulation. (b) A spin qubit in a double quantum dot: A magnetic field gradient $B_z(x)$ across the double-dot axis can be used to couple the electron spin and charge (orbital) degrees of freedom.  The electron spin can then couple to the resonator via an interaction of the form given in Eq.~\eqref{left-dot-coupling}.}
    \label{fig:double-dot}
\end{figure}

In this section, we derive the form of the longitudinal coupling between a microwave cavity constructed from a superconducting stripline resonator and a qubit encoded in either the charge or spin states of an electron in a double quantum dot (Fig.~\ref{fig:double-dot}). In each case (charge or spin qubits), we find the conditions under which the coupling can be modulated for only one of the qubit states [giving a form $\propto g_1(t)\ket{1}\bra{1}$ with $\delta g_0(t)=0$], as required for the which-path entangler.

The interaction between the charge density operator $\rho(\boldsymbol{r})$ for a qubit and the electric potential $V(\boldsymbol{r},t)$ due to a resonator can be written as
\begin{equation}\label{eq:Hrho}
    H_\rho = \int d^3 r \rho(\mathbf{r})V(\mathbf{r},t).
\end{equation}
When the potential $V(\mathbf{r},t)$ is a slowly-varying function of $\boldsymbol{r}$ on the scale of the qubit charge distribution, it can be expanded about the location $\mathbf{r}_0$ of the qubit, giving the electric dipole approximation (in the length gauge):
\begin{equation}\label{eq:DipoleExpansion}
H_\rho \simeq QV(\mathbf{r}_0,t)-\boldsymbol{d}\cdot\boldsymbol{E}(\mathbf{r}_0,t).
\end{equation}
Here, $\boldsymbol{E}(\mathbf{r}_0,t)=-\boldsymbol{\nabla}V(\mathbf{r}_0,t)$ is the cavity electric field at the position of the qubit. The quantities $Q$ and $\boldsymbol{d}$ are the total charge and dipole operators, respectively:
\begin{equation}
Q = \int d^3 r \rho(\boldsymbol{r});\quad \boldsymbol{d}=\int d^3 r \rho(\boldsymbol{r})(\boldsymbol{r}-\boldsymbol{r}_0).
\end{equation}
For qubit states constructed from systems with overall charge neutrality (e.g., neutral atoms and excitons), $Q\simeq 0$. An effective qubit-cavity interaction can then be derived by quantizing the cavity field and forming matrix elements of the dipole operator, as described in the main text and in Sec.~\ref{sec:Hamiltonian}, above. Since the dipole operator is odd under inversion about $\boldsymbol{r}_0$ ($\boldsymbol{d}\to -\boldsymbol{d}$), the longitudinal coupling will vanish for these systems (within the dipole approximation) if the qubit eigenstates $\ket{s}$ have definite parity under inversion. This will be the case for high-symmetry free-atom eigenstates, but it will not generally be true for lower-symmetry designed systems like the quantum dots considered here.

\subsubsection{Charge qubit in a double quantum dot}

In contrast to the far-field approach described above, many experiments that couple quantum dots to superconducting resonators instead exploit the strong near-field coupling of a quantum-dot charge to a metallic gate that extends from the resonator to the vicinity of the quantum dot \cite{mi2017strong,landig2018coherent,samkharadze2018strong} (Fig.~\ref{fig:double-dot}). For these systems, it is important to work directly from Eq.~\eqref{eq:Hrho} to find an accurate microscopic description of the coupling. To simplify the effective coupling, we now consider only a single quantized resonator mode. We further project the charge density operator onto the lowest two (left/right) orbital states of a double quantum dot. Thus, we approximate
\begin{eqnarray}
V(\boldsymbol{r},t) & \simeq & i\phi_0(\boldsymbol{r}) V_0(a^\dagger e^{i\omega_\mathrm{c}t}-ae^{-i\omega_\mathrm{c}t}), \label{eq:Vr}\\
\rho(\boldsymbol{r},t) & \simeq & -e|\psi_\mathrm{L}(\boldsymbol{r},t)|^2\ket{\mathrm{L}}\bra{\mathrm{L}}-e|\psi_\mathrm{R}(\boldsymbol{r},t)|^2\ket{\mathrm{R}}\bra{\mathrm{R}},\label{eq:rhort}
\end{eqnarray}
where the dimensionless mode function $\phi_0(\boldsymbol{r})$ solves Poisson's equation subject to the device geometry. Here, $V_0$ is the amplitude of zero-point voltage fluctuations in the resonator itself, and $\psi_l(\boldsymbol{r},t)$ ($l=\mathrm{L,R}$) is the envelope function for quantum dot $l$. We assume that the quantum-dot orbitals can be manipulated via external gate voltages, leading to adiabatically varying time-dependent envelope functions. This has been achieved in experiments on shuttling electrons between quantum dots \cite{seidler2022conveyor}. 

The coupling to the right dot $\ket{\mathrm{R}}$ is negligible if $\phi_0(\boldsymbol{r})$ is vanishingly small wherever the envelope function $\psi_\mathrm{R}(\boldsymbol{r},t)$ has significant weight. In this limit, we insert Eqs.~\eqref{eq:Vr} and \eqref{eq:rhort} into Eq.~\eqref{eq:Hrho}, giving the coupling between a double-dot charge qubit and a microwave resonator \cite{childress2004mesoscopic},
\begin{equation}
e^{-i\omega_\mathrm{c}a^\dagger a t} H_\rho e^{i\omega_\mathrm{c}a^\dagger a t} \simeq H_{\mathrm{int}}=ig_{c}(t)\ket{\mathrm{L}}\bra{\mathrm{L}}(a^\dagger-a),\quad g_{c}(t)=-e\alpha_{\mathrm{L}}(t)V_{0}.\label{left-dot-coupling}
\end{equation}
The coupling $g_c(t)$ between the charge and the resonator can thus be controlled through the lever arm $\alpha_\mathrm{L}(t)$, which can itself be controlled via the left dot orbital:  
\begin{eqnarray}
\alpha_\mathrm{L}(t) & = & \int d^3 r \phi_0(\boldsymbol{r})|\psi_\mathrm{L}(\boldsymbol{r},t)|^2,\label{eq:alphaL}\\
\alpha_\mathrm{R}(t) & = & \int d^3 r \phi_0(\boldsymbol{r})|\psi_\mathrm{R}(\boldsymbol{r},t)|^2\simeq 0.\label{eq:alphaR}
\end{eqnarray}
By manipulating the shape, size, and position of the left quantum dot via gate voltages, $\alpha_\mathrm{L}$ can therefore be modulated. The lever arm $\alpha_\mathrm{L}$ can equivalently be written in terms of the capacitance $C_\mathrm{L}$ between the left dot and the resonator [Fig.~\ref{fig:double-dot}(a)] and  the total capacitance $C_\Sigma$ of the left dot \cite{childress2004mesoscopic}:
\begin{equation}
\alpha_\mathrm{L} = \frac{C_\mathrm{L}}{C_\Sigma}.
\end{equation}
 
For a charge qubit encoded in the left/right basis ($\ket{\mathrm{L}}\to\ket{1}$, $\ket{\mathrm{R}}\to\ket{0}$), Eq.~\eqref{left-dot-coupling} directly achieves the form of longitudinal coupling required for the which-path entangler, where $g_c(t)\to g_1(t)$, and where $g_0(t)\simeq 0$ provided the cross-capacitance between the resonator and the right dot is negligible [equivalently, provided Eq.~\eqref{eq:alphaR} is satisfied]. Charge qubits of this type are sensitive to electric field fluctuations, leading to coherence times typically in the range of $T_2^*\sim 100\,\mathrm{ps}$-$10\,\mathrm{ns}$ \cite{petersson2010quantum}. These short coherence times would likely limit the applicability of pure charge qubits to the entangler scheme presented in the main text. For this reason, in the next section we consider longitudinal coupling for a spin qubit, since spin qubits typically have longer coherence times.

\subsubsection{Spin qubit in a double quantum dot}

An interaction of the form given in Eq.~\eqref{left-dot-coupling} can also be used to generate coupling between a microwave resonator and an electron spin. This generally requires coupling of the spin and charge degrees-of-freedom, since direct magnetic coupling of the electron spin magnetic moment to the resonator magnetic field is typically weak, on the order of a few tens of Hz~\cite{schoelkopf2008wiring,imamouglu2009cavity}. In the absence of intrinsic spin-orbit coupling, a synthetic spin-orbit interaction can be generated from a magnetic field gradient across the two dots~\cite{burkard2023semiconductor} [Fig.~\ref{fig:double-dot}(b)]. For a magnetic field oriented along the $z$ axis, the Hamiltonian describing spin-charge hybridization is given by~\cite{beaudoin2016coupling}
\begin{equation}
    H_{\mathrm{q}}(t)=\frac{1}{2}[\varepsilon\tau_z+\Omega(t) \tau_x]+\frac{1}{2}(b_{\mathrm{L}}\ket{\mathrm{L}}\bra{\mathrm{L}}+b_{\mathrm{R}}\ket{\mathrm{R}}\bra{\mathrm{R}})\sigma_z,\label{flopping-mode}
\end{equation}
where $\sigma_z$ is the Pauli-Z operator of the spin, $\varepsilon$ is the double-dot detuning controlling the relative dot potentials [Fig.~\ref{fig:double-dot}(b)], and where $b_{\mathrm{L,R}}=g^*\mu_{\mathrm{B}}\int d^3r \lvert \psi_{\mathrm{L,R}}(\bm{r})\rvert^2 B_z(\bm{r})$. Here, $g^*$ is the (material-dependent) electron $g$-factor and $\mu_{\mathrm{B}}$ is the Bohr magneton.  The tunnel splitting $\Omega(t)$ can be controlled through a gate voltage and will be used to modulate the coupling. In Eq.~\eqref{flopping-mode}, we have also introduced the orbital pseudospin Pauli operators
\begin{eqnarray}
\tau_x & = & \ket{\mathrm{L}}\bra{\mathrm{R}}+\ket{\mathrm{R}}\bra{\mathrm{L}},\\
\tau_z & = & \ket{\mathrm{R}}\bra{\mathrm{R}}-\ket{\mathrm{L}}\bra{\mathrm{L}}.
\end{eqnarray}

Equation~\eqref{flopping-mode} can be diagonalized conditioned on the spin state $\ket{\sigma}$ (where $\sigma=\uparrow,\downarrow$ with $\sigma_z\ket{\sigma}=\pm \ket{\sigma}$), giving 
\begin{align}
    \begin{aligned}
        &\ket{+,\sigma}_t=\mathrm{cos}\frac{\theta_\sigma(t)}{2}\ket{\mathrm{R}}\ket{\sigma}+\mathrm{sin}\frac{\theta_\sigma(t)}{2}\ket{\mathrm{L}}\ket{\sigma},\\
        &\ket{-,\sigma}_t=-\mathrm{sin}\frac{\theta_\sigma(t)}{2}\ket{\mathrm{R}}\ket{\sigma}+\mathrm{cos}\frac{\theta_\sigma(t)}{2}\ket{\mathrm{L}}\ket{\sigma},
    \end{aligned}
\end{align}
where $\mathrm{tan}\:\theta_{\uparrow,\downarrow}(t) =\Omega(t)/[\varepsilon\pm \Delta b_z]$ for $\Delta b_z=(b_{\mathrm{R}}-b_{\mathrm{L}})/2$. When the Zeeman energy is small compared to the double-dot orbital energy, the lowest two energy eigenstates are predominantly distinguished by spin $\sigma=\uparrow,\downarrow$, but they will also have slightly different charge distributions. The two states $\ket{-,\sigma}_t$ then encode the qubit:
\begin{eqnarray}
    \ket{0(t)} & = & \ket{-,\uparrow}_t,\\
    \ket{1(t)} & = & \ket{-,\downarrow}_t.
\end{eqnarray}
In order to obtain an effective qubit-resonator interaction, we now project Eq.~\eqref{left-dot-coupling} (with a time-independent $g_{c}$) into the qubit subspace  and transform to the instantaneous adiabatic eigenbasis via the unitary $U=\sum\ket{s}\bra{s(t)}$ (as described in Sec.~\ref{sec:Hamiltonian}), giving
\begin{equation}
    H_{\mathrm{int}}^{\mathrm{eff}}(t)=UP(t) H_{\mathrm{int}}P(t)^\dagger U^\dagger=ig_0(t)\ket{0}\bra{0}(a^\dagger-a)+ig_1(t)\ket{1}\bra{1}(a^\dagger-a),\quad P(t)=\sum_\sigma \ket{-,\sigma}_t\bra{-,\sigma},
\end{equation}
where
\begin{align}
    \begin{aligned}
        &g_0(t)=g_c\mathrm{cos}^2\frac{\theta_\uparrow(t)}{2}=\frac{g_c}{2}\left(1+\frac{\varepsilon+\Delta b_z}{\sqrt{(\varepsilon+\Delta b_z)^2+\Omega^2(t)}}\right),\\
        &g_1(t)=g_c\mathrm{cos}^2\frac{\theta_\downarrow(t)}{2}=\frac{g_c}{2}\left(1+\frac{\varepsilon-\Delta b_z}{\sqrt{(\varepsilon-\Delta b_z)^2+\Omega^2(t)}}\right).
    \end{aligned}
\end{align}
For $\varepsilon=-\Delta b_z$, we then have $g_0=g_c/2$ (time-independent) for all finite values of the tunnel splitting $\Omega(t)$. Hence for this choice of parameters ($\varepsilon=-\Delta b_z$), modulating $\Omega(t)$ only affects the coupling of the resonator to state $\ket{1}$:
\begin{align}
    &g_0=\frac{g_c}{2},\\
    &g_1(t)=\frac{g_c}{2}\left(1-\frac{2\Delta b_z}{\sqrt{4\Delta b_z^2+\Omega^2(t)}}\right).
\end{align}
For $\Omega(t)=\bar{\Omega}+\delta \Omega(t)$ with $\bar{\Omega}\gg \lvert\Delta b_z\rvert, \lvert \delta\Omega(t)\rvert$, we then have $g_1(t)=\bar{g}_1+\delta g_1(t)$, where
\begin{equation}
    \delta g_1(t)\simeq \frac{g_c \Delta b_z}{\bar{\Omega}^2}\delta \Omega(t).
\end{equation}

\section{Finite-bandwidth corrections to idealized QWP states}

In this section, we quantify finite-bandwidth effects due to a finite duration of the input wavepacket $u(t)$. These corrections lead to imperfect transmission and reflection of the incident wavepacket for a qubit prepared in the state $\ket{0}$, leading to deviations from the idealized QWP states considered in the main text. 

For a qubit initialized in $\ket{+}=(\ket{\mathrm{1}}+\ket{\mathrm{0}})/\sqrt{2}$, the QWP state generated for an incident coherent state with  amplitude $\alpha_0$ is given by Eq.~\eqref{which-path} of the main text (here we consider the case $\xi=0$ since dephasing does not affect the present discussion):
\begin{equation}
    \ket{\Psi}=\frac{1}{\sqrt{2}}\left(\ket{1,\psi_1}+\ket{0,\psi_0}\right),\label{S:which-path}
\end{equation}
where for $s=0,1$,
\begin{equation}
    \ket{\psi_s}=\prod_{i=1,2}D_i(\alpha_{is})\ket{\text{vac}}.
\end{equation}
Here, $\ket{\text{vac}}$ is the vacuum, $D_i(\alpha)=e^{\alpha a_{v_i}^\dagger-\mathrm{h.c.}}$ is a displacement operator that generates a coherent state with amplitude $\alpha$ in mode $a_{v_i}$, and 
\begin{equation}
\alpha_{is}=(-1)^{i-1}\alpha_0\int_0^\infty dt\:\lvert v_{i}(t)\rvert^2=(-1)^{i-1}\alpha_0\int\frac{d\omega}{2\pi}\:\lvert v_i(\omega)\rvert^2\label{S:amplitude}   
\end{equation}
is the qubit-state-dependent stationary amplitude of the coherent state in mode $a_{v_i}$. The   reflected ($v_{1}$) and transmitted ($v_{2}$) pulses are given by $v_1(\omega)=R(\omega)u(\omega)$ and $v_2(\omega)=T(\omega)u(\omega)$, respectively, where $R(\omega)$ and $T(\omega)$ are the reflection and transmission coefficients, and where $u(\omega)$ is the waveform of the input pulse. The input pulse is assumed to have a duration $\sim\tau$.

Up to corrections that vanish for $\kappa\tau \rightarrow\infty$,  only one of $\alpha_{is}$ is nonzero for each value of $s$, provided
\begin{equation}
    \kappa_1=\kappa_2=\frac{\kappa}{2}.
\end{equation}
With this choice of $\kappa_{1,2}$, Eq.~\eqref{S:which-path} describes a coherent state whose which-path degree-of-freedom is entangled with the qubit: For a qubit in state $\ket{1}$, the incident coherent state is fully reflected, whereas for a qubit in state $\ket{0}$, it is fully transmitted, up to corrections that vanish for $\kappa\tau\rightarrow\infty$.  In order to quantify these corrections, we consider the fidelity $F=\lvert\braket{\Psi_\infty\vert\Psi_{\kappa\tau}}\rvert^2$  of the idealized QWP state $\ket{\Psi_\infty}$ having ideal which-path character, obtained by taking $\kappa\tau\rightarrow\infty$, relative to the QWP state $\ket{\Psi_{\kappa\tau}}$ obtained for finite $\kappa\tau$. This is easily accomplished using the relation $R(\omega)-1=T(\omega)$ that follows from input-output theory after setting $\kappa_1=\kappa_2=\kappa/2$, together with Eq.~\eqref{Ts} of the main text:
\begin{equation}
T(\omega)=(1-s)\frac{\kappa/2}{i\omega-\kappa/2}.
    \label{S:Ts}
\end{equation}
From Eq.~\eqref{S:amplitude}, we then have $\alpha_{11}=\alpha_0$, $\alpha_{21}=0$, as well as the non-trivial coherent-state amplitudes
\begin{align}
    &\alpha_{10}=\alpha_0\int\frac{d\omega}{2\pi}\lvert u(\omega)\rvert^2 \times \frac{\omega^2}{\omega^2+\kappa^2/4},\label{S:alpha10}\\
    &\alpha_{20}=-\alpha_0\int \frac{d\omega}{2\pi}\lvert u(\omega)\rvert^2\times\frac{\kappa^2/4}{\omega^2+\kappa^2/4},\label{S:alpha20}
\end{align}
which depend on $u(\omega)$. Since the state $\ket{1,\psi_1}$ appearing in Eq.~\eqref{S:which-path} describes a perfectly reflected wavepacket for any $\kappa\tau$, the fidelity $F$ is controlled entirely by $\ket{0,\psi_0}$, where ideally (i.e.~for $\kappa\tau\rightarrow\infty$), $\ket{\psi_0}=D_2(\alpha_0)\ket{\text{vac}}$. The fidelity is therefore given by
\begin{equation}
    F=\frac{1}{4}\bigg\lvert1+e^{-\tfrac{1}{2}\lvert \alpha_{0}\rvert^2}e^{\alpha_0^*\alpha_{20}}\prod_{i=1,2}e^{-\frac{1}{2}\lvert \alpha_{i0}\rvert^2}\bigg\rvert^2,
\end{equation}
where $\alpha_{10}$ and $\alpha_{20}$ are given by Eqs.~\eqref{S:alpha10} and \eqref{S:alpha20}. As an example, we consider an input waveform $u(\omega)$ having a Gaussian envelope with width $\sim 1/\tau$:
\begin{equation}
    \lvert u(\omega)\rvert^2=2\sqrt{\pi}\tau e^{-\omega^2\tau^2}.
\end{equation}
Here, the normalization of $u(\omega)$ is set by the requirement that $\int dt\lvert u(t)\rvert^2=1$. For this choice of $u(\omega)$, Eqs.~\eqref{S:alpha10} and \eqref{S:alpha20} give $\alpha_{10}=\alpha_0(1-\sqrt{\pi}ye^{y^2}\mathrm{erfc}\:y)$ and $\alpha_{20}=-\alpha_0\sqrt{\pi}ye^{y^2}\mathrm{erfc}\:y$, where $y=\kappa\tau/2$. Taylor expanding in $(\kappa\tau)^{-1}\ll 1$, we find that $\alpha_{10}\simeq \alpha_0[2(\kappa\tau)^{-2}-12(\kappa\tau)^{-4}]$ and $\alpha_{20}\simeq -\alpha_0[1-2(\kappa\tau)^{-2}+12(\kappa\tau)^{-4}]$, giving 
\begin{equation}
    F=\frac{1}{4}\left(1+e^{-4\frac{\lvert \alpha_0\rvert^2}{(\kappa\tau)^4}+\mathrm{h.o.t.'s}}\right)^2\simeq 1-4\frac{\lvert \alpha_0\rvert^2}{(\kappa\tau)^4},
\end{equation}
where $\mathrm{h.o.t.'s}$ designates terms that are higher order in $1/\kappa\tau$.   For a Gaussian wavepacket, neglecting the subleading corrections therefore requires that $N=\lvert \alpha_0\rvert^2\ll(\kappa\tau)^4$. Throughout the rest of this supplement, we neglect corrections due to finite $\kappa\tau$ and consider only the limit $\kappa\tau\rightarrow\infty$.

\section{Two-qubit concurrence}

In this section, we derive a formula for the concurrence accounting for pure-dephasing noise on the control qubit, photon loss in the transmission lines, and detection errors arising from both imperfect coherent-state distinguishability and imperfect detector efficiency. The result derived here corresponds to Eq.~\eqref{concurrence} of the main text.

Starting from the QWP state [cf.~Eq.~\eqref{which-path} of the main text]
\begin{equation}
    \ket{\Psi_{\xi(t)}}=\frac{1}{\sqrt{2}}(e^{\frac{i}{2}\theta_\xi(t)}\ket{1,\psi_1}+e^{-\frac{i}{2}\theta_\xi(t)}\ket{0,\psi_0}),\label{S:QWP-noise}
\end{equation}
where $\theta_\xi(t)=\int_0^t dt'\xi(t')$ is a noise-dependent random phase, we model the effect of photon loss by inserting a fictitious beamsplitter into each interferometer arm. These beamsplitters are modeled by the unitary operator~\cite{zhang2013quantumS}
\begin{equation}
    B_{c,c'}(\varphi)=e^{i\frac{\varphi}{2}(c^\dagger c'+\mathrm{h.c.})} ,
\end{equation}
which describes scattering from mode $c$ into loss mode $c'$ with probability $p=\mathrm{cos}^2(\varphi/2)$. Although the loss from the two arms may be asymmetric in reality (with probability $p_i$ for loss from arm $i$), the expression for the concurrence derived here provides a lower bound for the concurrence achieved with asymmetric losses, if we take $p=\mathrm{max}\{p_1,p_2\}$. (Increasing the incoherent loss rate in either arm can only decrease entanglement.) 

Acting on Eq.~\eqref{S:QWP-noise} with $\prod_{i=1,2}B_{a_{v_i},a_{v_{i}}'}(\varphi_i)$, then tracing out modes $a_{v_{1}}'$ and $a_{v_{2}}'$ gives the reduced density matrix [still conditioned on a single realization of $\xi(t)$]
\begin{equation}
    \varrho_{\xi}(t)=\frac{1}{2}\left[\ket{\Phi_1}\bra{\Phi_1}+\ket{\Phi_0}\bra{\Phi_0}+e^{-pN}(e^{i\theta_\xi(t)}\ket{\Phi_1}\bra{\Phi_0}+\mathrm{h.c.})\right],\label{S:rhophi}
\end{equation}
where $N=\lvert \alpha_0\rvert^2$, and where we have introduced
\begin{align}
    \ket{\Phi_1}=D_1(\alpha_0\sqrt{1-p})\ket{1,\text{vac}},\quad\ket{\Phi_0}=D_2(-\alpha_0\sqrt{1-p})\ket{0,\text{vac}}.\label{S:basis}
\end{align}
Photon loss therefore leads to a suppression of the off-diagonal terms in Eq.~\eqref{S:rhophi} (dephasing). This effect is accompanied by a reduction of the coherent-state amplitude: $\alpha_0\rightarrow\alpha_0\sqrt{1-p}$. Note that our ability to write $\ket{\Phi_0}$ in terms of a single displacement operator is a consequence of neglecting finite-$\kappa\tau$ corrections.

In order to re-encode the which-path degree-of-freedom in a parity degree-of-freedom, the coherent states comprising the which-path qubit are interfered at a beamsplitter (described by unitary $U_{\mathrm{BS}}$) that transforms the output modes $a_{v_i}$ into new modes
$a_\pm=(a_{v_1}\pm a_{v_2})/\sqrt{2}$. Following this operation, the resulting field in the $a_-$ mode is unentangled with either the qubit or $a_+$ mode. Tracing out the $a_-$ mode ($\mathrm{Tr}_{a_-}\{\}$) and averaging ($\llangle\rrangle$) over realizations of the noise $\xi(t)$ then gives the following reduced density matrix for the state of the qubit and $a_+$ mode: 
\begin{equation}
    \varrho(t)=\llangle \mathrm{Tr}_{a_-}\{U_{\mathrm{BS}}\varrho_{\xi}(t)U_{\mathrm{BS}}^\dagger\}\rrangle =\frac{1}{2}\sum_{s=0,1}\ket{\Psi_{\sigma}}\bra{\Psi_{\sigma}}+\frac{1}{2}e^{-pN-\chi_\xi(t)}\left(\ket{\Psi_1}\bra{\Psi_0}+\mathrm{h.c.}\right),\label{entangled}
\end{equation}
where $\ket{\Psi_0}=\ket{0,-\alpha_p}$ and $\ket{\Psi_1}=\ket{1,+\alpha_p}$ for $\alpha_p=\alpha_0[(1-p)/2]^{1/2}$, and where for zero-mean, Gaussian, stationary noise having spectral density $S(\omega)=\int dt\:e^{-i\omega t}\llangle \xi(t)\xi\rrangle$,
\begin{equation}
    \chi_\xi(t)=\int\frac{d\omega}{2\pi}\frac{4\mathrm{sin}^2\left(\tfrac{\omega t}{2}\right)}{\omega^2}S(\omega).
\end{equation}
The photonic degree-of-freedom can then be entangled with a second qubit, initialized in $\ket{+}$, through an interaction that imparts a phase shift sending $\ket{+}\rightarrow\ket{-}$, conditioned on the electric field having an odd photon-number parity~\cite{besse2018singleS,kono2018quantumS,hacker2019deterministicS,besse2020parityS}. The photon-number parity can be identified by re-expressing $\ket{\pm \alpha_p}$ in terms of the even- and odd-parity cat states $\ket{C_\pm}=N_\pm(\ket{+\alpha_p}\pm \ket{-\alpha_p})$, where $N_\pm=(2\pm 2e^{-2\lvert \alpha_p\rvert^2})^{-1/2}$.   In addition to the approaches presented in Refs.~\onlinecite{besse2018singleS,kono2018quantumS,hacker2019deterministicS,besse2020parityS}, the required parity-conditioned phase flip could also be implemented by reflecting the incoming field off a single-sided cavity dispersively coupled to a qubit, i.e.,~coupled through an interaction of the form $\chi \sigma_z a^\dagger a$. In that case, the reflection coefficient (where $\omega$ is relative to the bare cavity frequency) is given by
\begin{equation}
    R(\omega)=\frac{i(z\chi-\omega)-\kappa/2}{i(z\chi-\omega)+\kappa/2},
\end{equation}
where $z=+ 1$ ($z=-1$) for a qubit in state $\ket{1}$ ($\ket{0}$). In this setup, an input coherent state occupying a spatiotemporal mode of duration $\tau$, resonant with $\chi$ (corresponding to the cavity frequency conditioned on the qubit being in state $\ket{1}$), will acquire a $\pi$ phase shift (sending $\ket{\pm \alpha}\rightarrow\ket{\mp \alpha}$) provided $\kappa\tau$ is large. With the qubit in state $\ket{0}$, however, and provided $2\lvert\chi\rvert\gg\kappa$, where $\kappa$ is the cavity decay rate (requiring that $\chi\gg\kappa\gg1/\tau$), the same field will be far off resonance and will therefore be reflected without a phase shift. 

Homodyne detection of the electric field quadrature along the axis of coherent-state displacement can be described by the two-element positive operator-valued measure (POVM) $P_\pm=(1\pm C_\theta)/2$, where $\theta$ (not to be confused with $\theta_\xi$, defined above) is the phase of $\alpha_p=e^{i\theta}\lvert \alpha_p\rvert$, and where~\cite{dragan2001homodyneS}
\begin{equation}
    C_\theta=\int_0^\infty dx\: H_\theta(x)-\int_{-\infty}^0 dx\:H_\theta(x),\quad 
    H_\theta(x)=\frac{1}{\sqrt{\pi(1-\eta)}}\mathrm{exp}\bigg\{-\frac{(x/\sqrt{\eta}-\hat{x}_\theta)^2}{1/\eta-1}\bigg\}.\label{S:quadrature}
\end{equation}
Here, $\eta$ is the detection efficiency and $\hat{x}_\theta=(e^{i\theta}a_+^\dagger+e^{-i\theta}a_+)/\sqrt{2}$. The operator $C_\theta$ [Eq.~\eqref{S:quadrature}] has the following symmetries with respect to $\ket{\pm \alpha_p}$~\cite{dragan2001homodyneS}: 
\begin{align}
    &\braket{\alpha_p\lvert C_\theta\rvert \alpha_p}=-\braket{-\alpha_p\lvert C_\theta\rvert {-}\alpha_p}=\mathrm{erf}(\sqrt{2\eta}\lvert \alpha_p\rvert)\label{S:symm1}\\
    &\braket{\alpha_p\lvert C_\theta\rvert {-}\alpha_p}=\braket{-\alpha_p\lvert C_\theta\rvert \alpha_p}=0.\label{S:symm2}
\end{align} 

We define Krauss operators $K_\pm$ given by $K_\pm = U\sqrt{1\pm C_\theta}$, where $U$ is a unitary. This can be done since $\lvert\lvert C_\theta\rvert\rvert\leq 1$. The post-measurement state of the qubits can then be written as
\begin{equation}
    \Tilde{\varrho}_\pm(t)=\frac{\mathrm{Tr}_{a_+}\{K_\pm \varrho(t) K_\pm^\dagger\}}{\mathrm{Tr}\{K_\pm^\dagger K_\pm  \varrho(t)\}},
\end{equation}
where $\mathrm{Tr}_{a_+}\{\dotsm\}$ denotes a partial trace over the state of the $a_+$ mode. Using the symmetries of $C_\theta$ given in Eqs.~\eqref{S:symm1} and \eqref{S:symm2}, we then find that
\begin{equation}
     \tilde{\varrho}_{\pm}(t)=p_\pm\varrho_{+}(t)+p_\mp\varrho_{-}(t),\label{S:bellstates}
 \end{equation}
where $p_\pm=\frac{1}{2}[1\pm \mathrm{erf}(\sqrt{2\eta}\lvert\alpha_p\rvert)]$, and where the states $\varrho_\pm(t)$ are given by  $2\varrho_\pm(t) =(1\pm \sigma_z\sigma_z)+e^{-pN-\chi_\xi(t)}(\sigma_x\sigma_x\mp \sigma_y\sigma_y)$. In the computational basis $\{\ket{0,0},\ket{0,1},\ket{1,0},\ket{1,1}\}$, these can be written in matrix form as
\begin{equation}
    \varrho_+(t)=\frac{1}{2}\begin{pmatrix}
    1 & 0 & 0 & e^{-pN-\chi_\xi(t)}\\ 
    0 & 0 & 0 & 0\\
    0 & 0 & 0 & 0\\
    e^{-pN-\chi_\xi(t)} & 0 & 0 & 1
    \end{pmatrix},\quad \varrho_-(t)=\frac{1}{2}\begin{pmatrix}
    0 & 0 & 0 & 0\\
    0 & 1 & e^{-pN-\chi_\xi(t)} & 0\\
    0 & e^{-pN-\chi_\xi(t)} & 1 & 0\\
    0 & 0 & 0 & 0
    \end{pmatrix}.\label{S:X1}
\end{equation}
Note that in the absence of both dephasing and photon loss ($p=\xi=0$), the above states are Bell states: $\varrho_\pm=(\ket{+,+}\pm \ket{-,-})/\sqrt{2}$ for $\ket{\pm}=(\ket{1}\pm \ket{0})/\sqrt{2}$.
\begin{figure}
    \centering
    \includegraphics[width=0.5\textwidth]{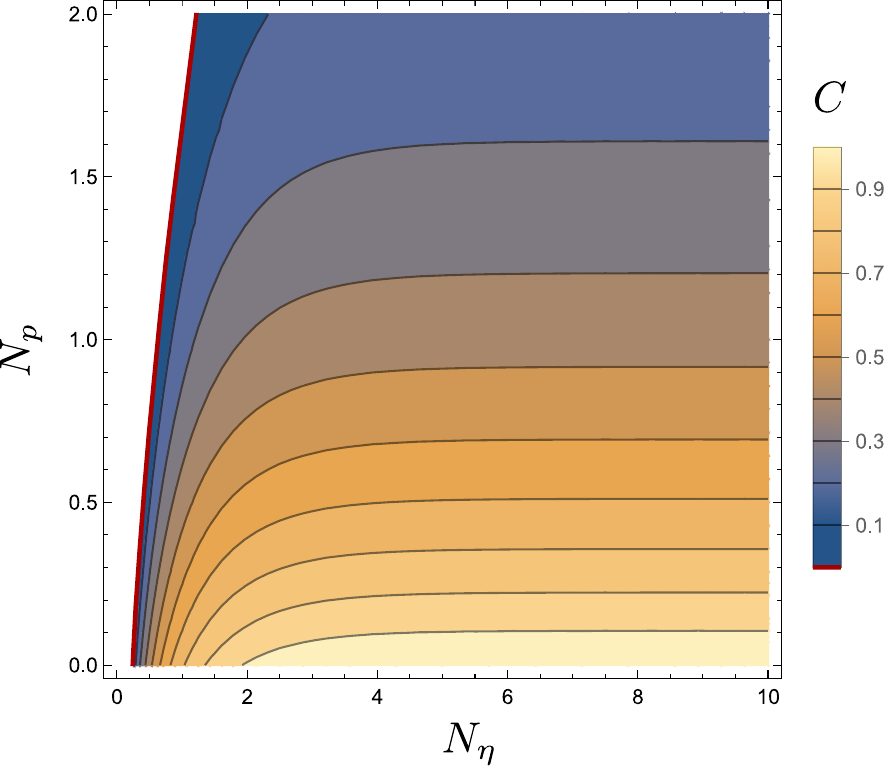}
    \caption{Sudden death of entanglement: For each value of $N_{\eta}=\eta(1-p)N$, there is a critical threshold in $N_{p}=pN$ beyond which the two-qubit concurrence vanishes. The contour $C=0$ is indicated in red.}
    \label{fig:concurrence}
\end{figure}
\begin{figure}
    \centering
    \includegraphics[width=0.5\textwidth]{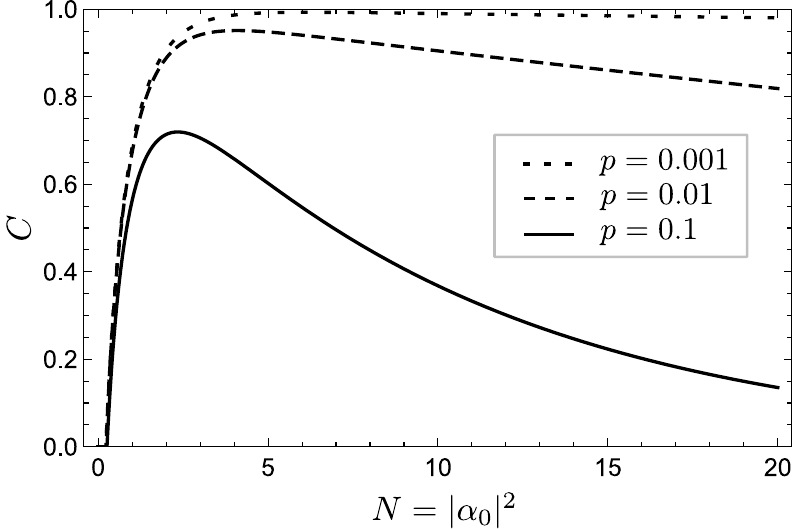}
    \caption{The concurrence [Eq.~\eqref{S:concurrence}] with $\xi(t)=0$ and $\eta=1$ for $p=0.001$ (dotted line), $p=0.01$ (dashed line), and $p=0.1$ (solid line). }
    \label{fig:concurrence-fixed-loss}
\end{figure}

Given Eq.~\eqref{S:X1}, it is easily seen that $\Tilde{\varrho}_\pm(t)$ is an X state~\cite{yu2005evolutionS,yu2006quantumS}, so called because its nonzero entries form an `X' shape. For an X state of the form
\begin{equation}
    \begin{pmatrix}
        a & 0 & 0 & w\\
        0 & b & z & 0\\
        0 & z^* & c & 0\\
        w^* & 0 & 0 & d
    \end{pmatrix},
\end{equation}
where $a+b+c+d=1$, the concurrence $C$~\cite{hill1997entanglementS, wootters1998entanglementS, wootters2001entanglementS} is easy to calculate~\cite{yu2005evolutionS}: $C=2\mathrm{max}\{0,\lvert z\rvert-\sqrt{ad},\lvert w\rvert-\sqrt{bc}\}$. For $\Tilde{\varrho}_\pm(t)$, this gives $C(t)=\mathrm{max}\{0,(1-\delta)e^{-pN-\chi_\xi(t)}-\delta\}$, where $\delta=p_-=\mathrm{erfc}(\sqrt{2\eta}\lvert\alpha_p\rvert)/2$. In terms of $N_{\eta}=\eta(1-p)N$ and $N_p=pN$, this result can be written as
\begin{equation}
    \boxed{C(t)=\mathrm{max}\{0,\mathrm{erf}(\sqrt{N_{\eta}})e^{-N_{p}-\chi_\xi(t)}-\mathrm{erfc}(\sqrt{N_{\eta}})\}.}\label{S:concurrence}
\end{equation}
This recovers Eq.~\eqref{concurrence} of the main text. We plot the concurrence as a function of $N_{\eta}$ and $N_{p}$ with $\xi(t)=0$ (Fig.~\ref{fig:concurrence}) in order to illustrate the sudden death of entanglement. In Fig.~\ref{fig:concurrence-fixed-loss}, we plot $C$ as a function of $N$ for $\eta=1$ in order to illustrate that for any fixed value of $p$, there exists a value of $N$ such that the concurrence is maximized.

\providecommand{\noopsort}[1]{}\providecommand{\singleletter}[1]{#1}%

\end{document}